\documentstyle[twoside,fleqn,npb,epsfig]{article}

\newcommand{\epem}{\mbox{$\mathrm{e}^+\mathrm{e}^-$}}
\newcommand{\as}{\mbox{$\alpha_s$}}

\newcommand{\AmS}{{\protect\the\textfont2
  A\kern-.1667em\lower.5ex\hbox{M}\kern-.125emS}}

\title{Event Shapes and Power Corrections in \epem\ Annihilations}

\author{G.~Dissertori\address{EP Division, CERN, \\
        CH-1211 Geneva 23, Switzerland}}

\begin{document}

\begin{abstract}
  \vspace*{-52mm} \hbox{ 
Talk presented at the DIS99 Workshop, Zeuthen/Berlin,
Germany, April 19-23, 1999.}
  \vspace*{48mm} 
Hadronic final states in \epem\ annihilations  at centre-of-mass energies from
14 GeV up to 189 GeV are studied in order to test recent predictions for
power corrections to the mean values as well as the distributions
of event shape variables.
\end{abstract}

% typeset front matter (including abstract)
\maketitle

%%%%%%%%%%%%%%%%%%%%%%%%%%%%%%%%%%%
% Introduction
%%%%%%%%%%%%%%%%%%%%%%%%%%%%%%%%%%%
\section{Introduction}
 \label{intro}

When studying observables in the process 
$\epem \rightarrow \mathrm{Hadrons}$, it is
found that the perturbative QCD predictions
have to be complemented by non-perturbative
corrections of the form $1/Q^p$, where 
$Q$ is the centre-of-mass energy $E_{CM}$,
and the power $p$ depends on the particular
observable. For fully inclusive observables
such as the total cross section
$p=4$, however, for less inclusive ones
such as event shape variables the power is much smaller,
typically $p=1$. So for those observables non-perturbative
effects can be sizeable, e.g., at LEP1
energies corrections of 5-10\% are found. Since
these variables are extensively used for \as\
determinations, non-perturbative effects have to
be well understood. Until recently they have been
determined from QCD-inspired Monte Carlo (MC) models of
hadronization, however, this method leads to
model dependence and thus limitations
in the precision of the \as\ measurements. The
new approach of power law corrections to event
shapes, pioneered by Dokshitzer and Webber \cite{Webber1}, 
could lead to improvements in this respect.

The event shape variables studied are Thrust, 
C-parameter, Heavy Jet Mass and Total and Wide Jet
Broadening.
% and the differential 2-jet rate with the
%Durham clustering algorithm. 
These are infrared
and collinear safe variables, and perturbative
predictions are known  up to second order in
\as, as well as the resummations of leading and
next-to-leading logarithms to all orders.

The results presented in the following are mostly
based on the study of Ref.~\cite{Hasko}, where
data from \epem\ annihilations at $E_{CM}=14$ GeV
up to 161 GeV have been analyzed. In addition,
some preliminary results from the LEP2 runs up to
189 GeV have been employed.

%%%%%%%%%%%%%%%%%%%%%%%%%%%%%%%%%%%
% Power corrections
%%%%%%%%%%%%%%%%%%%%%%%%%%%%%%%%%%%
\section{Power Corrections}
 \label{powers}

Power corrections are supposed to have their origin in
infrared divergences (renormalons) in the perturbative
expansions when the
overall energy scale $Q$ approaches the Landau pole
$\Lambda$. The first approaches were based on the
assumption of the existence of a universal
non-singular behaviour of an effective strong 
coupling at small scales, parametrized by a non-perturbative
parameter
\begin{equation} 
  \alpha_0(\mu_I) = \frac{1}{\mu_I} \int_{0}^{\mu_I} dk\; \as(k) \; ,
\end{equation}
with $\Lambda \ll \mu_I \ll Q$, which separates
the perturbative from the non-perturbative region, $\mu_I =2$ GeV,
typically. $\alpha_0(\mu_I)$ is assumed to be universal.

%%%%%%%%%%%%%%%%%%%%%%%%%%%%%%%%%%%
% mean values 
%%%%%%%%%%%%%%%%%%%%%%%%%%%%%%%%%%%
\subsection{Mean Values}
 \label{mean}

Using the Ansatz described above, the following prediction is
obtained for the mean value  of an event shape variable $f$~:
\begin{equation}
  \langle\, f\, \rangle = \langle\, f\, \rangle^{pert} \;+\; \langle\, f\, \rangle^{pow} \; ,
\end{equation}
where $\langle\, f\, \rangle^{pert}$ is the full second order prediction of the form
\begin{equation} 
            \as(\mu^2) A_f + 
            \as^2(\mu^2) \left[ B_f + A_f b_0 \ln\frac{\mu^2}{s} \right] \; .  
\end{equation}
Here $\mu^2$ is the renormalization scale, $s=E^2_{CM}$, and $b_0 = (33-2n_f)/(12\pi)$,
$n_f$ being the number of active flavours.

The power correction term is given by $\langle\, f\, \rangle^{pow} = a_f {\cal P}$,
where $a_f$ is 2 for Thrust, 1 for Heavy Jet Mass and $3\pi$ in case of the
C-parameter. $\cal P$ is a universal function of the form
%\begin{equation}
$
  {\cal P}  \approx \;{\cal M}\; \mu_I\; \alpha_0(\mu_I) / Q \;
$
%\end{equation}
(up to a constant and corrections of order \as\ and $\as^2$).
The {\it Milan factor} ${\cal M} \approx 1.8$ \cite{milanfactor} 
takes into account two-loop effects. Recently it has been found 
\cite{newbroad} that
in the case of the Jet Broadening variable the power correction is of a more 
complicated type compared to above, namely of the form $1/(Q \sqrt{\as(Q)})$.    

\begin{figure}[htb]
 \begin{center}
  \vspace{-1cm}    
  \includegraphics[width=7cm]{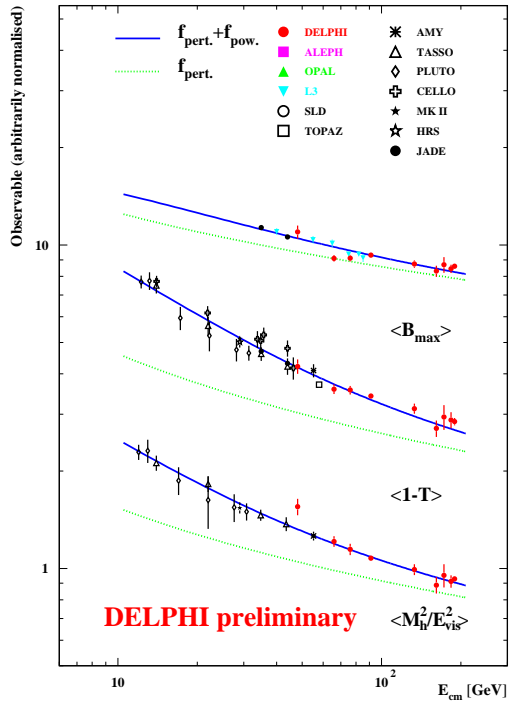}
  \vspace{-1cm}    
  \caption{Mean values of event shape variables as a function of the
           centre-of-mass energy.
           \label{fig:delphimeans}}
 \end{center}  
 \vspace{-1cm}
\end{figure}

DELPHI \cite{delphi} have measured mean values for Thrust, 
Wide Jet Broadening and
Heavy Jet Mass from the LEP1 and LEP2 data and combined 
their results with measurements from
 low energy \epem\ experiments in order to extract $\as(M_Z)$ and $\alpha_0$ from
a fit of the power law  Ansatz to these data. The fits are displayed in 
Fig.\ \ref{fig:delphimeans}. Very good fits are obtained with $\as(M_Z)$
between 0.118 and 0.120, and $\alpha_0(2 \mathrm{GeV})$ between 0.40 and 0.55. Similar results have been found in the analysis of Ref.~\cite{newbroad}.

%%%%%%%%%%%%%%%%%%%%%%%%%%%%%%%%%%%
% distributions
%%%%%%%%%%%%%%%%%%%%%%%%%%%%%%%%%%%
\subsection{Distributions}
 \label{distributions}

\begin{figure}[htb]
 \begin{center}
  \vspace{-1cm}    
  \includegraphics[width=7cm]{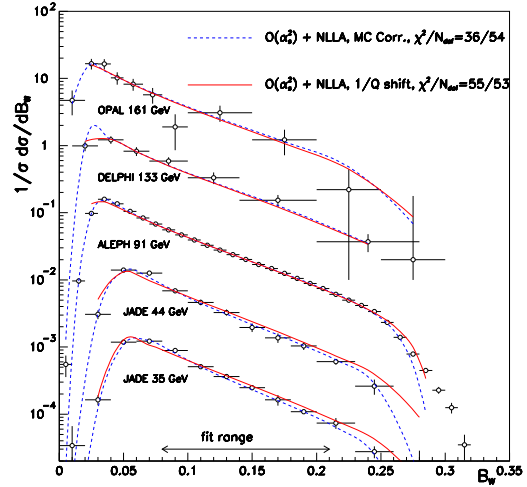}
  \vspace{-1cm}    
  \caption{Fits to distributions of the Wide Jet Broadening 
          at  several centre-of-mass energies. \label{fig:bw_dist}}
 \end{center}  
 \vspace{-1cm}
\end{figure}

For distributions of event shape observables it has been shown
\cite{dokdis} that the non-perturbative corrections lead to a shift
in the distribution, i.e,
\begin{equation}
 \frac{1}{\sigma_{tot}} \frac{d \sigma(f)^{corr}}{d f} = 
 \frac{1}{\sigma_{tot}} \frac{d \sigma(f - \Delta f)^{pert}}{d f}
\end{equation}
where in the cases of Thrust, Heavy Jet Mass and C-parameter 
the shift $\Delta f$
is given by exactly the same terms as the correction for the mean values, i.e.,
$\Delta f = a_f {\cal P}$. An improved calculation \cite{newbroad} for the Jet
Broadening variable has shown that in this case the distribution is not only
shifted, but also squeezed, since the shift is of the form 
$\Delta B \propto {\cal P} \ln(1/B)$.
The perturbative distribution is obtained from a matching of the 
full next-to-leading
order prediction to the  resummation of all leading and 
next-to-leading logarithms $\ln f$.
Theoretical uncertainties on the \as\ and $\alpha_0$ determinations 
are estimated
from variations of the renormalization scale and the scheme applied for 
matching
the fixed order and resummed calculations. Central values are given for 
$\mu^2=s$.
In Fig.\ \ref{fig:bw_dist} the fits to the Wide Jet Broadening are displayed, 
for various centre-of-mass energies. Good fits are obtained for the power law
Ansatz as well as for the more traditional approach of hadronization 
corrections from
MC models. Similar fits to Thrust and Heavy Jet Mass work well at 
high energies,
however, some deviations are found at very small energies. 
There probably the limit of 
applicability of the power law approach is reached. Furthermore, 
mass effects could play a role there.

\begin{figure}[htb]
 \begin{center}
  \vspace{-1cm}    
  \includegraphics[width=7cm]{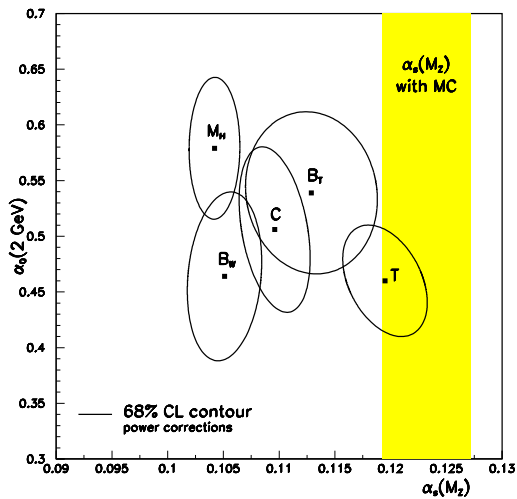}
  \vspace{-1cm}    
  \caption{Summary of the power law fits to distributions of event shapes.
           \label{fig:distsum}}
 \end{center}  
 \vspace{-1cm}
\end{figure}

In Fig.\ \ref{fig:distsum} a summary of the results can be found. 
A combination of  the results gives
$\as(M_Z)  =  0.1082 \pm 0.0021$, 
$\alpha_0(2 \mathrm{GeV})  =  0.504 \pm 0.042$. 
For $\alpha_0$ universality is found at the level of 20\%. 
The value of the
strong coupling results to be lower than the one obtained from 
a similar fit when using MC models
for the hadronization corrections, namely $\as(M_Z) = 0.1232 \pm 0.0040$.
This difference has still to be understood. 

%%%%%%%%%%%%%%%%%%%%%%%%%%%%%%%%%%%
% shape functions
%%%%%%%%%%%%%%%%%%%%%%%%%%%%%%%%%%%
\section{Non-Perturbative Shape Functions}
 \label{shapefunctions}

Recently it has been shown \cite{Sterman} that all leading
power corrections of the type $1/(fQ)$, f being the event shape variable, can
be resummed when folding the perturbative distribution with a
non-perturbative \textit{shape function}. 
%\begin{equation}
% \frac{1}{\sigma_{tot}} \frac{d \sigma(f)^{pert}}{d f} \rightarrow
%\int d\epsilon S(\epsilon)  \frac{1}{\sigma_{tot}} 
%\frac{d \sigma(f-\epsilon/Q)^{pert}}{d f} \; .
%\end{equation}
The form of the shape function depends on the observable, 
and new non-perturbative
parameters are introduced. Fits have been tried for 
Thrust and Heavy Jet Mass. For the former a good fit quality for a 
large energy range is obtained, and $\as(M_Z)$ values close to 
the world average are found. However, in case of the latter no satisfactory
fits could be achieved. This should be followed up in the future analyses. 

%\begin{figure}[htb]
% \begin{center}
%  \vspace{-1cm}    
%  \includegraphics[width=8cm]{thrust_shape.eps}
%  \vspace{-1cm}    
%  \caption{Relations between   vertices and colour factors
%           \label{fig:thrust_shape}.}
% \end{center}  
% \vspace{-1cm}
%\end{figure}

%%%%%%%%%%%%%%%%%%%%%%%%%%%%%%%%%%%
% conclusions
%%%%%%%%%%%%%%%%%%%%%%%%%%%%%%%%%%%
\section{Conclusions}
 \label{conclusions}

Significant progress has been made in the understanding
of power corrections to event shape variables in \epem\ annihilations.
Universality of the non-perturbative parameter $\alpha_0$ is observed
at the level of 20\%. Some open questions remain such as the difference
of \as\ values obtained with power laws and MC corrections. Also the 
effects of quark or hadron masses should be studied.
Power law predictions for other
variables such as the differential two-jet rate as well as the energy-energy
correlations are awaited for. 

A new approach based on non-perturbative shape functions looks very promising,
but some further investigations are required.

%%%%%%%%%%%%%%%%%%%%%%%%%%%%%%%%%%%
% thanx
%%%%%%%%%%%%%%%%%%%%%%%%%%%%%%%%%%%
\section{Acknowledgements}
 \label{thanx}

I would like to thank H.~Stenzel for providing me with the 
results of his power law studies, and G.~Salam and G.P.~Korchemsky
for helpful discussions.

%%%%%%%%%%%%%%%%%%%%%%%%%%%%%%%%%%%
% bibliography
%%%%%%%%%%%%%%%%%%%%%%%%%%%%%%%%%%%

\end{document}